# Control of surface and bulk crystalline quality in single crystal diamond grown by chemical vapour deposition

I. Friel[†], S. L. Clewes, H. K. Dhillon, N. Perkins, D. J. Twitchen, G. A. Scarsbrook

Element Six Ltd, King's Ride Park, Ascot, SL5 8BP, United Kingdom

**Abstract:**

In order to improve the performance of existing technologies based on single crystal diamond grown by chemical vapour deposition (CVD), and to open up new technologies in fields such as quantum computing or solid state and semiconductor disc lasers, control over surface and bulk crystalline quality is of great importance. Inductively coupled plasma (ICP) etching using an Ar/Cl gas mixture is demonstrated to remove sub-surface damage of mechanically processed surfaces, whilst maintaining macroscopic planarity and low roughness on a microscopic scale. Dislocations in high quality single crystal CVD diamond are shown to be reduced by using substrates with a combination of low surface damage and low densities of extended defects. Substrates engineered such that only a minority of defects intersect the epitaxial surface are also shown to lead to a reduction in dislocation density. Anisotropy in the birefringence of single crystal CVD diamond due to the preferential direction of dislocation propagation is reported. Ultra low birefringence plates (< $10^{-5}$) are now available for intra-cavity heat spreaders in solid state disc lasers, and the application is no longer limited by depolarisation losses. Birefringence of less than $5\times10^{-7}$ along a direction perpendicular to the CVD growth direction has been demonstrated in exceptionally high quality samples.

[†]Corresponding author

1.    Introduction

Chemical vapour deposition (CVD) growth of single crystal diamond has progressed rapidly in recent years, stimulating much research work and leading to the emergence of commercial technology based on this material. Electronic devices [1,2] and biological sensors [3], to name but two research developments, have been demonstrated, whilst CVD single crystal detectors [4,5], tool plates and optical windows are all now available commercially [6].

In order to increase the range of technological applications of single crystal CVD diamond it is important that further developments are made in four key areas which can be considered to define the state of the art of the material. These key areas are: size, purity, surface quality and crystal perfection, and to a greater or lesser extent these properties are important in all applications.

Nearly all technological applications of single crystal diamond, from cutting tool blades to die for the fabrication of transistors, will benefit from increases in the available plate sizes. CVD single crystal diamond is grown homoepitaxially, such that the basic size is to a large extent governed by the size of the substrate. The use of large natural diamond substrates is not a realistic option due to the prohibitive cost and limited availability. The most common starting templates for CVD growth are therefore high pressure high temperature (HPHT) type Ib substrates. At the present time sizes greater than about 6 × 6 mm$^2$, sufficiently free of defects to be suitable as substrates, are not routinely available although suitable plates up to 8 × 8 mm$^2$ are occasionally available. Thus it has long been desirable to develop methods based on CVD growth to increase the size of the crystal beyond the original substrate. Several methods have been reported which lead to increases in size such as the use of lateral growth beyond the substrate perimeter [7] or the growth of diamond layers in stages (including growth along orthogonal directions) [8,9,10]. The use of lateral growth appears promising but is limited by the defective or twinned material which generally accompanies such growth. Another option for producing large plates is by the method of tiling smaller plates [11]. However, although this route may have some applications they may be limited in utility by the defects which form due to the inevitable misorientation of adjacent plates. Increasing the size of single crystal CVD diamond plates commercially available therefore remains a challenge; the limit at this time being about 10 × 10 mm$^2$ [6].

Controlling the purity of single crystal CVD diamond is concerned with limiting the incorporation of unwanted elements, most commonly nitrogen and boron. Several groups have demonstrated material with excellent bulk electronic properties [12,13], and a prerequisite for this is the ability to reduce the point defect densities to the parts per billion level. A new challenge for diamond growers at the present time concerns controlling isotopic purity ($^{13}$C/$^{12}$C ratio) in addition to elemental purity for diamond quantum computing applications [14].

In contrast to many other established technological materials, such as silicon, surface processing of single crystal diamond often presents as great a challenge to the materials developer as that of controlling other properties required by the application, such as doping concentrations, material purity or extended defect densities. The ability to process surfaces in order to planarize, etch or form geometrical features opens up a whole range of possible applications. In a material as hard and as brittle as single crystal diamond it is extremely difficult to process surfaces mechanically without imparting sub-surface damage. In addition, although dry etching of diamond has been widely reported (see for example [15] and references therein), the ability to etch surfaces over large areas without preferentially etching damage or defects, and thus roughen the surface, presents a significant challenge. In Section 3 we report on methods developed to remove sub-surface damage whilst maintaining planarity on a macroscopic scale and simultaneously maintaining low roughness on the microscopic scale.

Crystal perfection in the context of high quality single crystal diamond pertains to controlling extended defects such as dislocations, rather than defects normally associated with polycrystalline growth such as grain boundaries, which are not formed in true homoepitaxial diamond growth. Several advanced applications of single crystal diamond require material containing a low density of extended defects, such as low birefringence diamond heat spreaders in high power lasers. In Section 4 we consider the types of extended defects commonly encountered in synthetic single crystal diamond and several methods we have used to reduce the extended defect densities in CVD diamond.

2. Experimental Methods

All samples reported here were grown by microwave plasma CVD in a $CH_4$ and $H_2$ gas atmosphere at growth temperatures in the range of 700°C - 950°C. Homoepitaxial growth was performed on {100}-oriented HPHT or CVD single crystal substrates typically 0.5 mm in thickness. Further details of the deposition conditions used are given in [16].

X-ray topography (XRT) was used to characterise the extended defect content of selected samples. Details of the methods used are given fully in [17]. The surface morphology of the diamond samples was characterised by differential interference contrast microscopy (DIC), and by atomic force microscopy (AFM) using a Digital Instruments Dimension 3100.

Quantitative birefringence microscopy was performed using the commercially available *Metripol* system [18]. The Metripol optics, which are retro-fitted onto a polarizing microscope, consist of a filter for producing monochromatic light (λ = 550 nm was used here), a rotatable linear polarizer, a quarter-wavelength retardation plate, a fixed linear analyser, and a CCD camera for image capture. The transmitted intensity, *I*, through the sample—optics system is given by

$$I = \tfrac{1}{2} I_0 \{1 + \sin[2(\alpha - \varphi)] \sin \delta \}, \qquad (1)$$

where $I_0$ is the incident intensity, α is the rotation angle of the linear polarizer, φ is the local angle of the slow axis (that is, the direction of polarisation corresponding to the local maximum in refractive index) and δ is the phase shift through the sample (of thickness *d*) between components along the fast and slow axes due to birefringence Δ*n*, given by

$$\delta = \frac{2\pi}{\lambda} (\Delta n)\, d. \qquad (2)$$

The transmitted intensity for a set of values of α is measured. A fitting procedure is then used to calculate |sin δ|. From this a two-dimensional false-colour map of |sin δ| is generated (from which Δ*n* can be determined). Metripol images reported here are restricted to samples for which $0 \leq \delta \leq \pi/2$ to avoid ambiguity.

For the purposes of defining what is meant by "ultra low birefringence diamond" it is useful to do so within the context of a real technological application. Here the application of a single crystal diamond plate as a cooling element within the resonant cavity of a high powered laser is considered (see Section 4). For linearly polarised light, during one round trip the light passes twice through the diamond window and back again through the polarizer. It can be shown [19] that the fraction of light transmitted back through the polarizer is given by

$$I/I_0 = 1 - \sin^2(2\varphi) \sin^2(\delta/2) \qquad (??)$$

The depolarisation loss can be estimated by simply averaging over all φ and assuming Δ*n* is approximately constant, such that $I/I_0 \approx \tfrac{1}{2}\sin^2(\delta/2)$ per round trip. For a typical plate thickness of 0.5 mm, a birefringence Δ*n* = $1\times10^{-5}$ leads to a loss of around 0.2 %. Clearly the loss is a function of the retardation (Δ*n*) × *d*, not Δ*n* alone. However, since Δ*n* is an indicator of crystalline quality independent of sample thickness, Δ*n* ≤ $1\times10^{-5}$ is adopted here as a definition of ultra low birefringence diamond.

## 3    Surface processing of single crystal diamond

## 3.1 Sub-surface damage due to mechanical processing

Low surface damage is a key requirement for most advanced technology based on single crystal diamond. As-grown CVD diamond surfaces are inherently free of processing damage. However, for many applications requiring growth of more than a few microns of diamond it can be difficult to avoid macroscopic features across the entire face of an as-grown layer such as growth hillocks [20], facets and macroscopic steps [21], and some form of bulk mechanical planarization step is usually required.

Due to diamond's brittle nature any mechanical process inevitably involves the removal of material from the surface by fracture using diamond grit, and this process can be expected to lead to surface micro-cracks [22]. As a general rule the larger the grit size used in polishing or lapping, the greater the extent of the damage (for a given polishing direction and plane). Although it is difficult to rigorously quantify sub-surface damage due to mechanical processing, it is possible to do so qualitatively by dry etching in, for example, an oxygen-containing plasma. This process removes damaged or defective material preferentially, leading to etch pits, the size and density of which can be used to compare levels of damage. However, the process also preferentially etches dislocations and other extended defects in the material such that for a highly defective sample it can be difficult to distinguish etch pits due to extended defects from those due to processing-induced damage. An example of two single crystal diamond samples etched using an oxygen-containing plasma and with differing levels of sub-surface damage are shown in Fig. 1.

## 3.2 Dry etching of single crystal diamond to remove sub-surface damage

The results shown in Fig. 1 also suggest that dry etching can be used to remove sub-surface damage from a mechanically processed surface. However if surface pitting and therefore surface roughening is to be avoided, the etch process should remove damaged areas, non-damaged areas and areas around extended defects at the same etch rate, ruling out the use of an oxygen-containing plasma. The use of inductively coupled plasma (ICP) etching of single crystal diamond using an Ar/Cl plasma chemistry was recently reported [23]. This plasma etching method was used to etch a mechanically-polished single crystal diamond substrate prepared with a deliberately-damaged surface. Fig. 2 shows a DIC micrograph of this surface post-Ar/Cl etching, compared to a similarly prepared substrate which underwent an oxygen-containing plasma etch. It can be seen that the Ar/Cl ICP etch does not etch damage preferentially, there being a complete absence of macroscopic features post-etch. In addition to this it is found that on a microscopic scale the surface does not roughen after Ar/Cl etching. Fig. 3 shows AFM images on a 1×1 µm$^2$ area of a low roughness, mechanically processed surface, before and after removal of approximately 500 nm of material using the Ar/Cl ICP etch.

### 3.3 Applications of single crystal diamond requiring low damage surfaces

Many applications of synthetic single crystal diamond should benefit from the ability to produce low damage surfaces. Substrates for single crystal CVD growth require low surface and sub-surface damage in order to minimise dislocations in the deposited layer (this is considered in detail in Section 4). Diamond monochromators for 3$^{rd}$ and 4$^{th}$ generation synchrotrons are intolerant to any strain in the crystal caused by sub-surface damage due to mechanical processing [24]. Applications involving optical centres in diamond, such as single photon sources where the light-emitting centre is in the near-surface region of the crystal, require that the surface is free from luminescence-quenching defects, which sub-surface damage is thought to contribute to [25].

In addition to etching bulk surfaces, ICP etching can also be used in conjunction with lithographic techniques to perform electronic device fabrication. The technique also opens up the possibility of forming geometrical structures in diamond for optical applications such as micro-lenses, gratings, waveguides or cavity resonators. Fig. 4 shows examples of spherical micro-lenses formed in single crystal diamond by Ar/Cl etching. Details of the lithography process used to form these structures are given in [26].

Such structures might be used in diamond-based quantum computing applications, in which low-damage micro-optical features formed by etching would be advantageous for efficient coupling of photons into and out of the optical centres of interest. Alternatively, Ar/Cl ICP etching could be used to "clean up" the surfaces of features formed using more aggressive processes such as focussed ion beam milling.

### 4. Reduction of extended defects in single crystal CVD diamond

In this section the types of extended defect commonly found in synthetic single crystal diamond are considered, with a focus on dislocations. Methods which can be used to reduce dislocation densities are considered in detail followed by an example of an application requiring low dislocation density single crystal diamond.

### 4.1 Types of extended defect in synthetic single crystal diamond

Twins are common planar defects occurring in diamond. The twinned crystal can be described by a rotation of the lattice through 180° about a <111> axis, thus forming a mirror image of the original crystal about the {111} plane [27]. Once formed twins are observed to increase in size both laterally and vertically as growth proceeds, leading to a macroscopic defect in the diamond crystal. Large twins can render the material unsuitable for further applications as they lead to local areas of degraded material properties, and the surrounding diamond lattice becomes highly strained. Since in diamond, twinning occurs on {111} planes and facets, for high quality homoepitaxial CVD growth on {100}-oriented substrates, twinning does not present a significant problem and is not considered further. Stacking faults are another planar defect occurring in single crystal diamond [27]. These defects are commonly observed in HPHT-grown material but, as with twin formation, only occur on {111} planes and are therefore not a common feature of {100}-oriented CVD growth.

More recently vacancy clusters have been proposed as the cause of the colouration in brown natural and some CVD diamond [28,29]. In CVD growth, such brown colouration is only observed when relatively high levels of nitrogen are incorporated into the solid (> 0.1 ppm), and in which the accelerated growth rates simultaneously lead to growth errors and hence a degradation in the material quality.

Once a basic CVD growth window for high quality diamond has been established [30] such that twins or other planar defects are not formed, the challenge in increasing the crystalline quality of the material becomes one of reducing the dislocation density.

4.2     Methods for reducing dislocations in single crystal CVD diamond

The two main sources of dislocations in single crystal CVD diamond are: (i) dislocations which nucleate at the interface between the substrate and the epitaxial layer due to errors in or disruptions to the lattice; (ii) dislocations or other extended defects which propagate from the substrate into the CVD layer. In either case these dislocations always propagate approximately parallel to the local CVD growth direction [21].

Dislocations nucleating at the epitaxial interface can be caused either by particulate contamination or by surface and sub-surface damage to the substrate. With careful cleaning and handling it is possible to reduce contamination to an acceptable level, such that controlling the sub-surface damage remains the major challenge (this was addressed in Section 3). A detailed X-ray topography study [17] of single crystal CVD diamond identified dislocations and dislocation clusters of the edge or 45° mixed type, with <110> Burgers vector. By deliberately growing on substrates with damaged surfaces and comparing to layers grown on carefully polished substrates these authors reported a clear increase in the density of 45° mixed dislocations on the damaged substrates. These mixed dislocations were found to have a component of their Burgers vector parallel to the polishing direction, further suggesting a correlation with surface damage.

Fig. 5 shows a set of {111} XRT projection topographs of three CVD layers of size approximately 3 × 3 × 2 mm$^3$. These plates were grown on type Ib HPHT substrates of similar quality (free of major cracks and inclusions and of low strain), under the same growth conditions, but with a differing quality of the substrate surface preparation. Sample A was grown on a highly damaged surface, achievable for example using a coarse diamond grit mechanical polishing process. Sample B was grown on a substrate with an intermediate level of sub-surface damage and sample C was grown on a substrate carefully prepared (using a cast iron scaif and plasma etch) to reduce damage.

It is clear from Fig. 5 that increasing the substrate surface quality leads to a dramatic reduction in the dislocation density. In Sample C it is believed that most of the dislocations present are those propagating through from the underlying substrate rather than being due to polishing damage, as evidenced by the fact that the dislocations do not nucleate in rows (e.g. along lines of polishing damage) but are randomly located.

Fig. 6 shows quantitative birefringence images of samples B and C for transmission parallel to the growth direction (sample A is not included as δ > π/2 for this sample). In agreement with the results of XRT there is a clear reduction in birefringence in the sample deposited on the substrate with the highest surface quality (sample C). For this viewing direction $d$ = 2 mm such that |sin δ| = 1 corresponds to Δ$n$ = 7×10$^{-5}$. Sample B has several areas in which the birefringence is at this level. Sample C has two local areas in which Δ$n$ = 4×10$^{-5}$ due to two large clusters of dislocations exhibiting the well-known petal-shaped birefringence pattern. On this basis it would appear that all three samples do not fall within the definition of ultra low birefringence diamond (Δ$n$ ≤ 1×10$^{-5}$). However, for light transmitted perpendicular to the growth direction it is observed that the birefringence is much lower (Fig. 7). When viewed along this direction (for which $d$ = 3 mm) sample A still contains some peripheral regions in which Δ$n$ > 1×10$^{-5}$. For sample B, Δ$n$ ≤ 6×10$^{-6}$ and for sample C, Δ$n$ ≤ 1×10$^{-6}$. This anisotropy in the birefringence is a characteristic of CVD diamond and is due to the preferred direction of propagation of the dislocations (along the growth direction). For light incident parallel to the growth direction, optical paths lying within the local strain field of a dislocation (or dislocation bundle) do so through the entire length of the sample, which maximises the phase shift between the components of polarisation along the fast and slow axes. On the other hand, light incident perpendicular to the growth direction, for low enough dislocation density, passes into and out of dislocation strain fields such that the overall phase shift is less. Ultra low birefringence diamond plates can therefore be produced through both a reduction in the dislocation density and also by processing the plate or designing the optical system in such a way that the optical path is perpendicular to the CVD growth direction. Processing should also be performed in such a way that no damage is imparted which might lead to strain.

By careful preparation of the substrate surface and also by selecting a substrate with a low defect density it is possible to reduce birefringence to less than $10^{-5}$ even along the growth direction, as shown in Fig. 8. Along the growth direction (for which $d$ = 3 mm), as shown in Fig. 8 (a), $\Delta n < 2 \times 10^{-6}$ over the entire area whilst perpendicular to the growth direction ($d$ = 3.3 mm), as shown in Fig. 8 (b), the birefringence is too low to be detected by the instrument and is background limited at $\Delta n < 5 \times 10^{-7}$.

In order to investigate defect propagation from the substrate, CVD growth was carried out on a type IIa HPHT substrate in which the epitaxial surface was prepared to minimise surface and sub-surface damage, and XRT performed with the substrate still attached. Fig. 9 shows three {111} projection topographs of the bi-layer structure. It can be seen that the IIa HPHT substrate has a relatively low dislocation density but does contain a series of planar defects around the periphery identified as stacking faults on {111} planes. Where these stacking faults intersect the epitaxial interface, rather than propagate continuously into the CVD layer, a row of dislocations is formed which subsequently propagate approximately parallel to the {100} growth direction. At present the mechanism by which these stacking faults in the substrate are converted into dislocations in the CVD layer is unclear. However, a careful examination of the sample indicates that the vast majority of defects in the CVD layer are due to those which propagate from the substrate. Furthermore, birefringence microscopy along the CVD growth direction reveals that $\Delta n < 1 \times 10^{-5}$ across the CVD plate, except within the strain fields of the dislocations nucleating above the stacking faults in the substrate. These results show that with careful substrate surface preparation it is possible to grow CVD diamond in which the extended defect density is limited by the substrate with little or no dislocations generated at the interface. Over the low-defect region (approximately 2 × 2 mm$^2$) the dislocation density is estimated to be < 400 cm$^{-2}$.

These results also suggest the possibility of reducing the dislocation density if the substrate can be engineered in such a way that the intersection of defects with the epitaxial surface is minimised. This can be achieved in substrates manufactured from single crystal CVD material by processing in such a way that the dislocation line directions (in other words the growth direction) lie in the plane of the substrate, ensuring only a small minority exit the epitaxial surface. To test this, a (001)-oriented single crystal CVD layer was grown on a substrate with a deliberately damaged surface in order to generate a high density of dislocations. A vertical plate was then extracted from this layer and processed into a second substrate with a (100)-oriented growth surface (orthogonal to the original growth surface). This substrate was processed in such a way as to minimise surface damage and hence the generation of new dislocations. CVD growth was then carried out along the [100] direction to produce a final block of CVD material containing two generations of CVD growth along the [001] and [100] directions. A {111} projection topograph of this block of material is shown in Fig. 10. The two orthogonal growth directions are clearly discernable from the dislocation lines in the topograph. Birefringence microscopy measured along the two growth directions yielded $\Delta n \leq 1\times 10^{-4}$ for the first stage [001] growth, and $\Delta n \leq 2\times 10^{-5}$ for the second stage [100] growth. This confirms that a reduction in the dislocation density is achievable using this method of sequential CVD growth along orthogonal directions, provided the surface for each growth stage is carefully prepared to minimise damage.

4.3     Applications of single crystal CVD diamond with reduced dislocation density

CVD diamond with reduced dislocation densities should improve the performance of several existing applications. For example, single crystal diamond power electronic devices should exhibit higher breakdown voltages. In addition, new applications in laser physics are now a possibility.

There has been much recent activity in the development of compact, medium and high power solid-state disc lasers and semiconductor-based vertical external cavity surface emitting lasers (VECSELs) [31]. Thermal management of the laser medium is a major challenge for developers of this technology, especially for applications requiring power scaling. Several workers have reported VECSEL designs in which a diamond heat-sink was attached to the back side of the semiconductor Bragg reflector, external to the laser cavity [32]. However, modelling confirms that thermal management is more efficient if the diamond heat sink is in contact with the output face of the semiconductor [33]. This method of heat removal is relatively independent of the thermal conductivity of the semiconductor. In conjunction with diamond's wide spectral transmittance, it is therefore an attractive thermal management scheme for a range of semiconductor lasers operating from the UV to the near IR. In this configuration, since the diamond sample is within the laser cavity, ultra low birefringence diamond is in general required in order to minimise depolarisation losses.

Miller et al [34] report a recent assessment of the depolarisation losses due to single crystal diamond used as an intra-cavity cooling element in a Nd:GdVO$_4$ disk laser. The experimental set up is shown schematically in Fig 11 (a), in which a Brewster plate was inserted within the cavity to enable a measurement of the polarisation-dependent losses to be made. Fig. 11 (b) shows the round trip loss as a function of the position of the beam for four single crystal diamond windows with different levels of birefringence. The use of ultra low birefringence synthetic diamond leads to much lower depolarisation losses compared to the windows made from natural diamond and single crystal CVD diamond with high birefringence. These results demonstrate that with the low levels of birefringence now achievable, single crystal synthetic diamond is now a viable option for thermal management in laser systems of this kind.

Relatively large area low birefringence plates were recently developed by growth on large HPHT Ib substrates, and through careful control of the crystal morphology. Fig. 12 shows a birefringence microscopy image of a 9 × 5 x 0.5 mm$^3$ plate, viewed perpendicular to the growth direction, with Δ$n$ = 8×10$^{-6}$ at the periphery and Δ$n$ < 5×10$^{-6}$ over the majority area of the plate. This increase in the available size of low birefringence plates, achieved by a combination of growth on a large area Ib HPHT substrate, and through careful control of the growth morphology, opens up the possibility of using diamond for intra-cavity thermal management in high powered solid state laser applications with larger beam diameters.

## 5      Summary

Inductively coupled plasma (ICP) etching using an Ar/Cl gas mixture has been demonstrated to remove sub-surface damage of mechanically processed surfaces, whilst maintaining macroscopic planarity and low roughness on a microscopic scale. Dislocations in high quality single crystal CVD diamond are shown to have been reduced by using substrates with a combination of low surface damage and low densities of extended defects. Substrates engineered such that only a minority of defects intersect the epitaxial surface are also shown to lead to a reduction in dislocation density. Anisotropy in the birefringence of single crystal CVD diamond due to the preferential direction of dislocation propagation has been reported. Ultra low birefringence plates (< 10$^{-5}$) for intra-cavity heat spreaders in solid state disc lasers are shown not to limit the application through depolarisation losses. Birefringence of less than 5×10$^{-7}$ along a direction perpendicular to the CVD growth direction has been demonstrated for exceptionally high quality samples.

These developments pave the way for enhanced performance of existing technologies based on single crystal CVD diamond, and have the potential to enable new technologies in fields such as quantum computing, electronics or laser physics.

## 6      Acknowledgements

Element Six Ltd would like to thank Mike Gaukroger and Phillip Martineau of DTC Research Centre, UK, for providing the X-ray topographs and many valuable discussions. We would also like to thank Chee-Leong Lee, Alan Kemp and Martin Dawson of the Institute of Photonics, University of Strathclyde, UK, for performing the ICP etching reported in Section 3 and for the use of the results of the solid state disc laser experiments in Section 4.

**8      Figures**

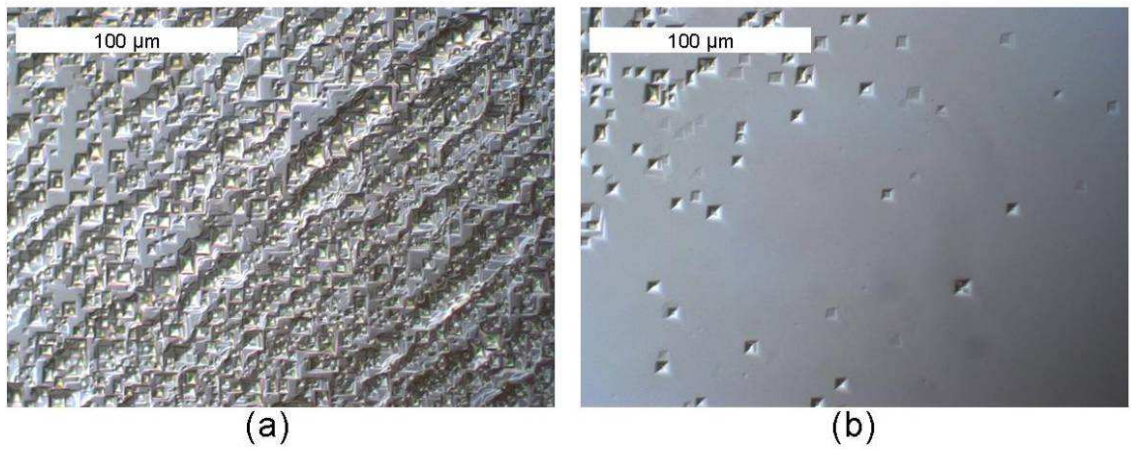

Fig. 1 – DIC micrographs of single crystal samples etched in an oxygen-containing plasma with (a) high levels and (b) lower levels of sub-surface damage.

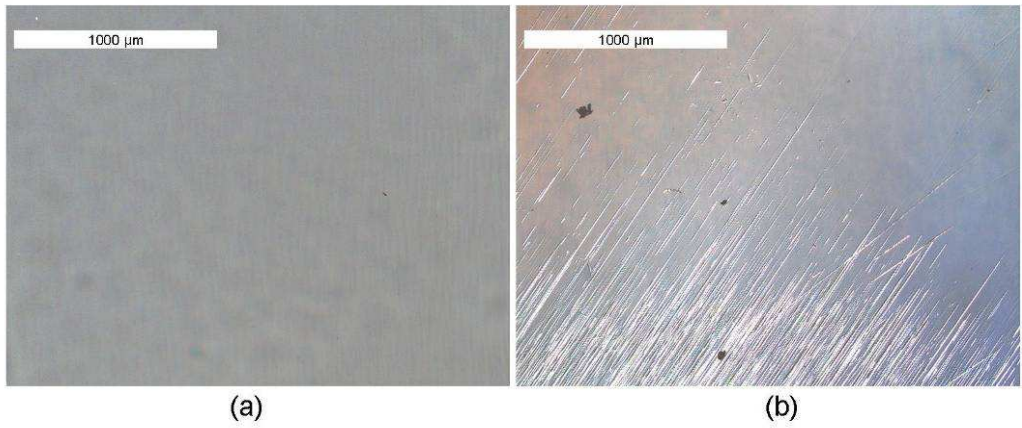

Fig. 2 - DIC micrograph of a single crystal diamond substrate etched using (a) Ar/Cl ICP and (b) an oxygen-containing plasma. Both substrates were mechanically polished using a relatively high-damage process.

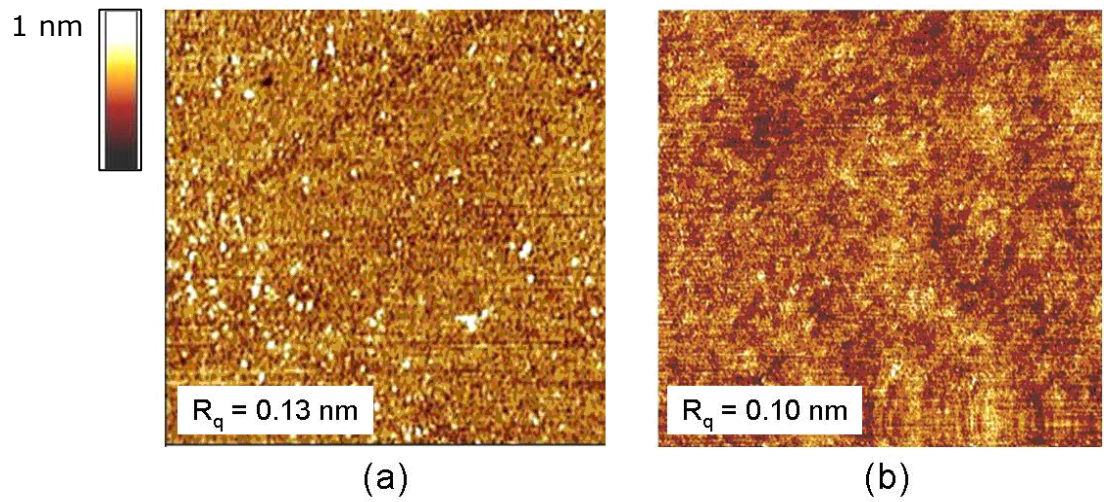

Fig. 3 - AFM images on a 1×1 µm² single crystal diamond surface (a) before and (b) after etching of ≈ 500 nm by Ar/Cl ICP.

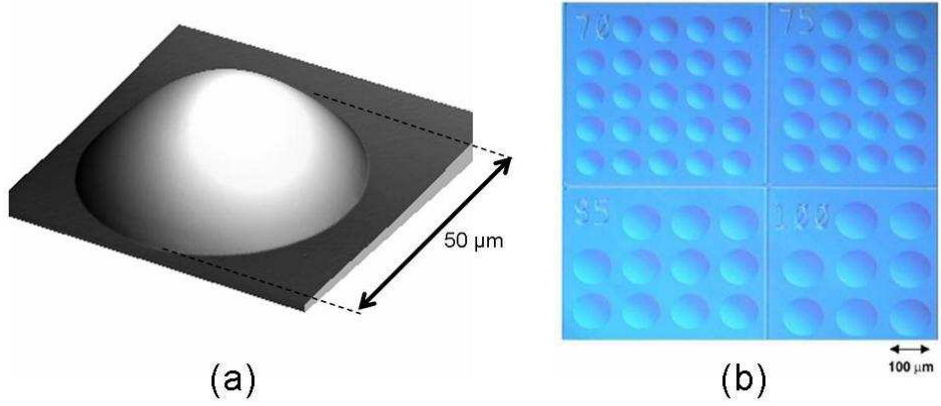

Fig. 4 – (a) AFM image of a spherical micro-lenses formed in single crystal diamond by Ar/Cl etching (the vertical height is approximately 800 nm); (b) optical image of an array of micro-lenses of various sizes.

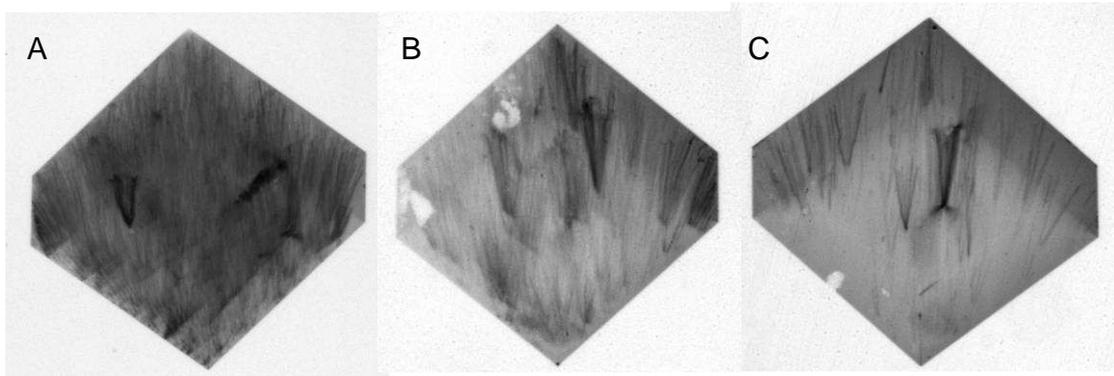

Fig. 5 – {111} projection topographs of three CVD layers grown on Ib HPHT substrates with high (sample A), medium (sample B) and low (sample C) levels of sub-surface damage.

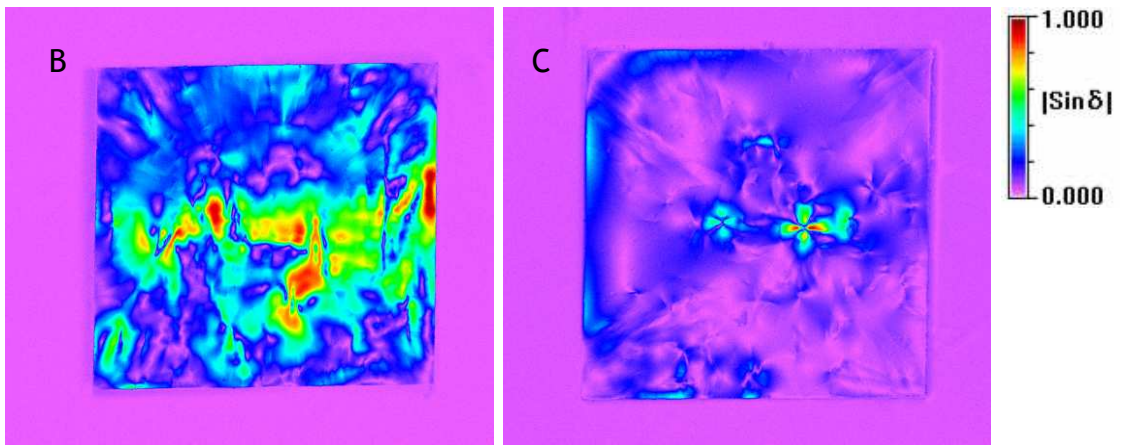

Fig. 6 - Quantitative birefringence images of samples B and C for transmission parallel to the growth direction.

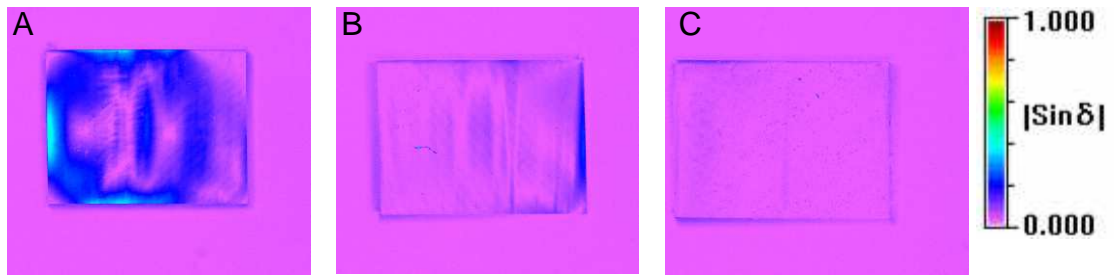

Fig. 7 - Quantitative birefringence images of samples A, B and C for transmission perpendicular to the growth direction.

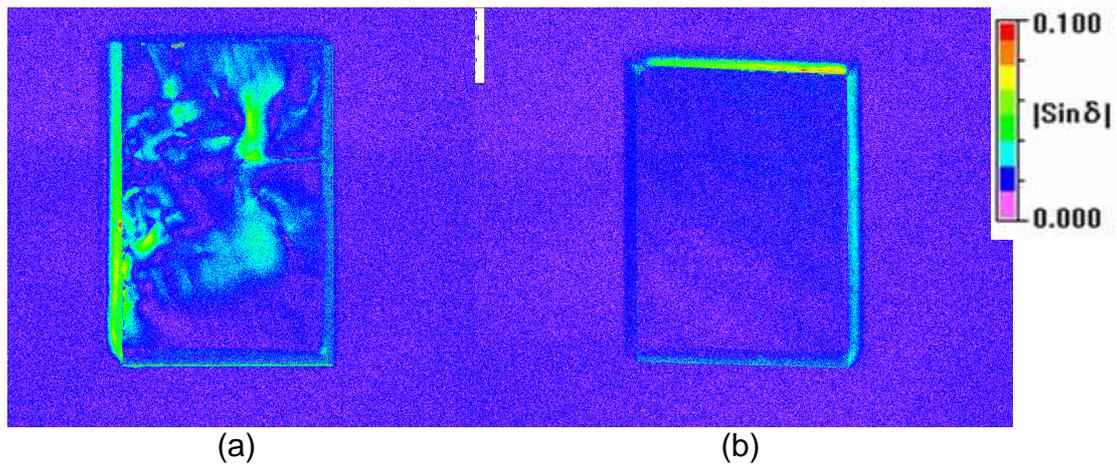

Fig. 8 - Quantitative birefringence images along (a) the growth direction and (b) perpendicular to the growth direction of a single crystal CVD sample deposited on a carefully prepared and selected type Ib substrate.

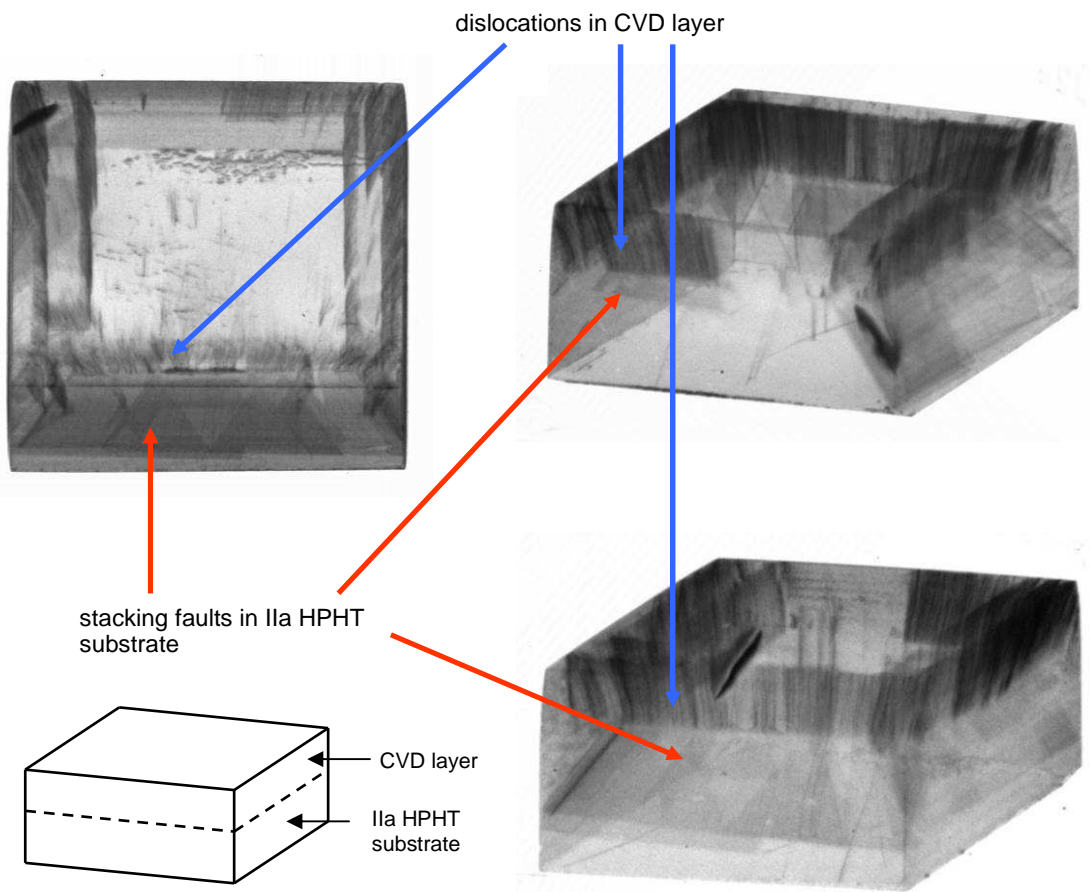

Fig. 9 - {111} projection topographs of a CVD layer deposited on a type IIa HPHT substrate. The approximate position of the epitaxial interface is indicated in the inset.

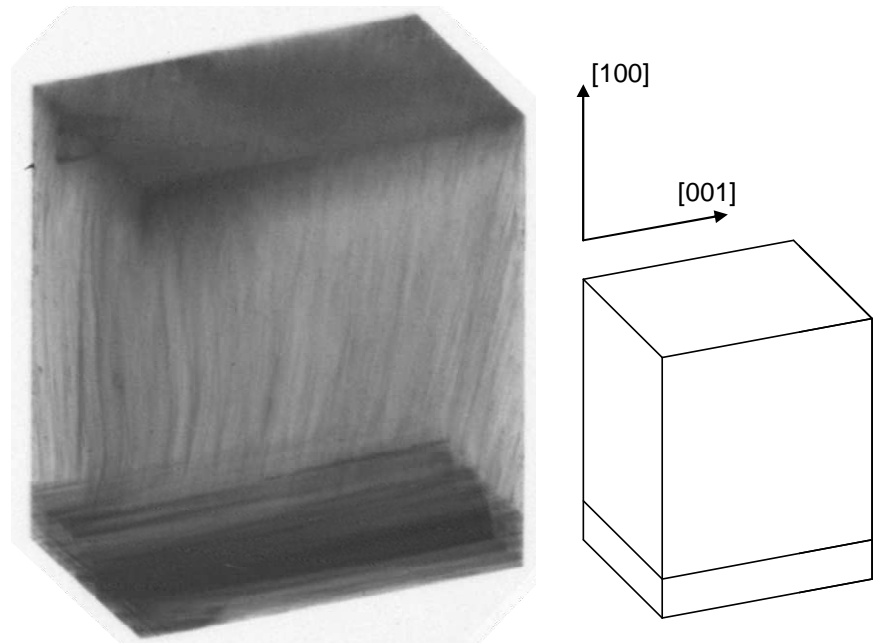

Fig. 10 – {111} projection topograph of a single crystal CVD diamond sample containing two generations of CVD growth along the [001] and [100] directions.

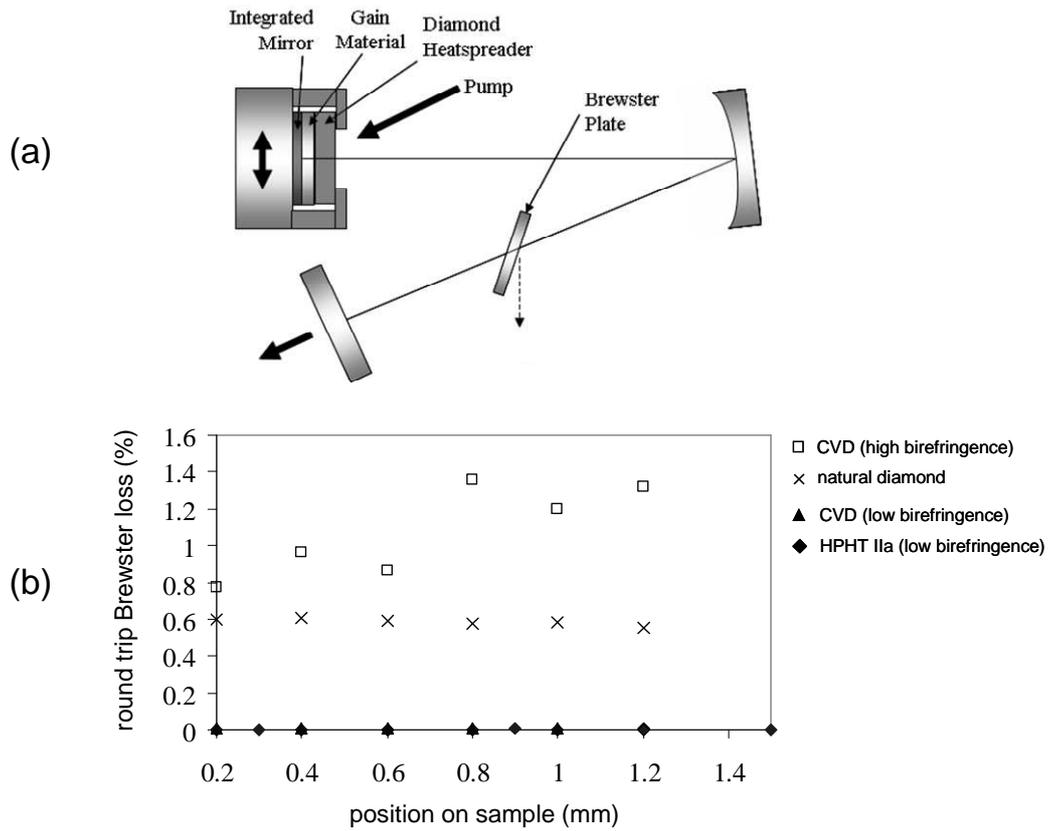

Fig. 11 – (a) experimental set up used to evaluate intra-cavity diamond heat spreaders in a Nd:GdVO$_4$ disk laser; (b) round trip loss as a function of the position of the beam on each diamond heat spreader. The low birefringence heat spreaders (closed symbols) exhibited Brewster losses of less than 0.01 %.

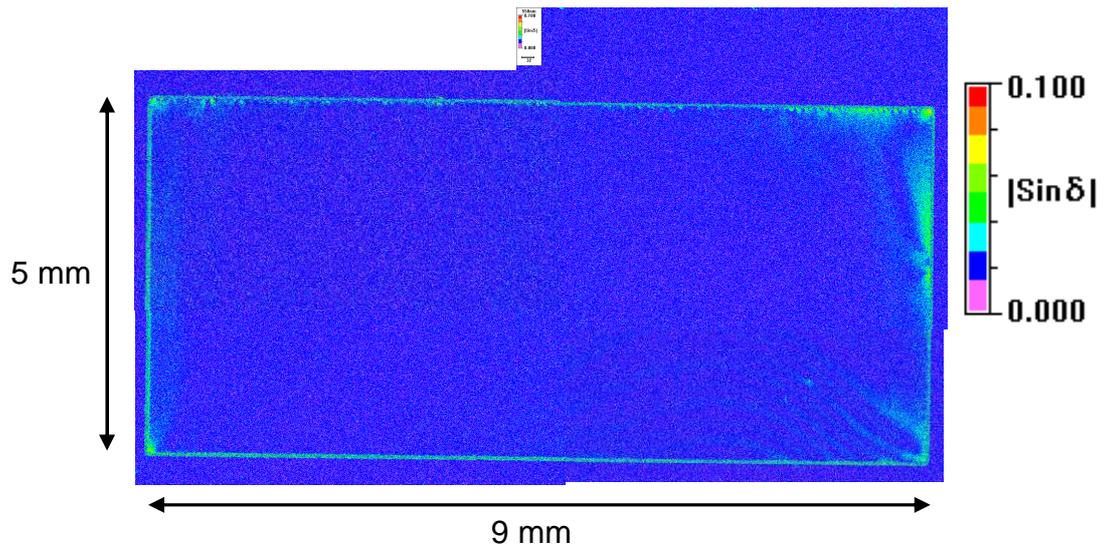

Fig. 12 - Birefringence microscopy image of a 9 × 5 x 0.5 mm³ plate (optical path perpendicular to the CVD growth direction).